\documentclass[a4paper,11pt]{article}
\pdfoutput=1 

\usepackage{jcappub} 

\usepackage{appendix}
\let\sb=_ \catcode`\_=\active \def_#1{\ensuremath \sb{\rm#1}}

\title{A closer look at interacting dark energy with statefinder hierarchy and growth rate of structure}

\author[a]{Jing-Lei Cui,}
\author[a]{Lu Yin,}
\author[a]{Ling-Feng Wang,}
\author[a]{Yun-He Li,}
\author[a,b,1]{Xin~Zhang\note{Corresponding author.}}

\affiliation[a]{Department of Physics, College of Sciences, Northeastern University, \\Shenyang
110004, China}
\affiliation[b]{Center for High Energy Physics, Peking University, \\Beijing 100080, China}

\emailAdd{owldasw@163.com}
\emailAdd{yinlumail@foxmail.com}
\emailAdd{qwert89526@163.com}
\emailAdd{liyh19881206@126.com}
\emailAdd{zhangxin@mail.neu.edu.cn}

\abstract{We investigate the interacting dark energy models by using the diagnostics of statefinder hierarchy and growth rate of structure. We wish to explore the deviations from $\Lambda$CDM and to differentiate possible degeneracies in the interacting dark energy models with the geometrical and structure growth diagnostics. We consider two interacting forms for the models, i.e., $Q_1=\beta H\rho_c$ and $Q_2=\beta H\rho_{de}$, with $\beta$ being the dimensionless coupling parameter. Our focus is the I$\Lambda$CDM model that is a one-parameter extension to $\Lambda$CDM by considering a direct coupling between the vacuum energy ($\Lambda$) and cold dark matter (CDM), with the only additional parameter $\beta$. But we begin with a more general case by considering the I$w$CDM model in which dark energy has a constant $w$ (equation-of-state parameter). For calculating the growth rate of structure, we employ the ``parametrized post-Friedmann'' theoretical framework for interacting dark energy to numerically obtain the $\epsilon(z)$ values for the models. We show that in both geometrical and structural diagnostics the impact of $w$ is much stronger than that of $\beta$ in the I$w$CDM model. We thus wish to have a closer look at the I$\Lambda$CDM model by combining the geometrical and structural diagnostics. We find that the evolutionary trajectories in the $S^{(1)}_3$--$\epsilon$ plane exhibit distinctive features and the departures from $\Lambda$CDM could be well evaluated, theoretically, indicating that the composite null diagnostic $\{S^{(1)}_3, \epsilon\}$ is a promising tool for investigating the interacting dark energy models. We also compare our results with the observed uncertainties on diagnostic parameters. We find that current observations still do not have sufficient precisions to completely distinguish I$\Lambda$CDM models from the $\Lambda$CDM model. Anyway, our work points out what precisions of measurements should be achieved to distinguish the I$\Lambda$CDM models from the $\Lambda$CDM model.}

\begin{document}
 \bibliographystyle{unsrt}
\maketitle


\section {Introduction}
\label{intro}
The cosmological observations~\cite{Riess:1998cb,Perlmutter:1998np,Tegmark:2003ud,Spergel:2003cb} have shown that our universe is undergoing accelerated expansion. This cosmic acceleration is attributed to an unknown component, called dark energy (DE), with negative pressure, or a modification to general relativity (GR) on the cosmological scale. For the possible existence of DE, a large number of DE models are constructed in theory, such as the $\Lambda$CDM model consisting of the cosmological constant ($\Lambda$) and the cold dark matter (CDM), the scalar-field models~\cite{quintessence,Ratra:1987rm,Caldwell:1999ew,Zhang:2005kj,Zhang:2005rg,Zhang:2006av,Sen:2002in,holotach,ArkaniHamed:2003uy,hologhost,ghost}, the holographic type models~\cite{Li:2004rb,Li:2012xf,Wei:2007ty,Gao:2007ep,holocyc,heal,nonflat}, and so on. On orientation of modified gravity, there are many attempts, for instance, the $f$(R) class theories~\cite{DeFelice:2010aj,Nojiri:2008nt}, the Dvali-Gabadadze-Porrati (DGP) braneworld model~\cite{Dvali:2000hr}, and the like. Among a lot of theories and models, the $\Lambda$CDM model is the simplest and provides a good fit to current observational data, though the other models cannot be completely excluded yet. Nevertheless, the $\Lambda$CDM model is not free of any troubles; it always suffers from the severe theoretical challenges, i.e., the fine tuning and coincidence problems~\cite{Sahni:1999gb,Padmanabhan:2002ji,Bean:2005ru,Copeland:2006wr,Sahni:2006rde,Kamionkowski:2007wv,limiao}.

Currently, in the empirical studies of cosmic acceleration, we are actually facing two physically profound questions: Is the cosmic acceleration caused by a breakdown of GR on cosmological scales or by a DE component that produces repulsive gravity within GR? If the cause of acceleration is indeed a DE component, is it a cosmological constant $\Lambda$ or a dynamical field? To differentiate DE and modified gravity (MG), the only usage of geometrical diagnostics is not enough, and the utilization of probes of structure growth is a must. In this paper, we do not consider the possibility of MG, but only focus on the studies of DE. But, actually, even if only the possibility of DE is considered, the combination of geometrical and structure growth diagnostics is proven to be very powerful and helpful in differentiating alternative models from $\Lambda$CDM, as will discussed in this paper.

Undoubtedly, any definitive deviation from $\Lambda$CDM, if diagnosed out and confirmed, would be a major breakthrough in cosmology and fundamental physics. One possible deviation comes from the models in which the cosmological constant $\Lambda$ is replaced by a dynamical dark energy. There exists, however, another possibility that DE (in the simplest case, the vacuum energy density or $\Lambda$) directly couples to CDM. The model of vacuum energy interacting, in some physically profound way, with CDM is called the I$\Lambda$CDM model in this paper. We will explore the deviation of this model from $\Lambda$CDM with the combination of geometrical and structure growth diagnostics in this work.

 To discriminate various models, the geometrical diagnostics are introduced, such as statefinder $\{r,~s\}$~\cite{Sahni:2002fz,Alam:2003sc}, $Om$, and $Om3$~\cite{om,Shafieloo:2012rs,Sahni:2014ooa}. Some works~\cite{sfide,sfhde,Zhang:2004gc,Zhang:2007uh,Setare:2007sf,Feng:2008sf,Tong:2009mu,Zhang:2009qa,Lu:2008hp,Cui:2014sma} revealed that the statefinder diagnostic can effectively discriminate various DE models and break/differentiate the degeneracies between different parameter values of DE model, except for a few models (for example, the new agegraphic DE model~\cite{Zhang:2009qa,Cui:2014sma}). Furthermore, Arabsalmani and Sahni~\cite{Arabsalmani:2011fz} introduced the statefinder hierarchy and the growth rate of linear perturbations as ``null diagnostics'' for the $\Lambda$CDM model. The statefinder hierarchy~\cite{Arabsalmani:2011fz} is also a geometrical diagnostic and is model-independent, which contains high derivatives of $a(t)$, with $a(t)$ the scale factor of the universe. The growth rate of structure~\cite{Acquaviva:2008qp,Acquaviva:2010vr} was presented previously as a scale-independent consistency check between the expansion history and the structure growth. It can act as a cosmic growth history diagnostic, or be combined with the statefinder hierarchy to serve on a composite diagnostic. In the previous work~\cite{Zhang:2014sqa}, we applied these two diagnostics to discriminate four holographic DE models. These diagnostics were also considered in~\cite{sfh1,Li:2014mua,Hu:2015opa}.

In this paper, we will use the statefinder hierarchy and the growth rate of structure to diagnose the deviation from the $\Lambda$CDM regarding the coupling between vacuum ($\Lambda$) and CDM in the I$\Lambda$CDM model. Though we pay more attention to the I$\Lambda$CDM model, we wish to begin with the more general model of interacting DE in which the equation of state (EoS) parameter $w$ is a constant and may not be exactly $-1$, usually referred to as the I$w$CDM model. We will first apply the geometrical and structural diagnostics in the I$w$CDM model, exploring and comparing the impacts of EoS $w$ and coupling $\beta$.

Consider a spatially flat Friedmann-Robertson-Walker (FRW) universe consisting of DE (de), CDM (c), baryons (b), and radiation (r). The energy conservation equations of cosmic components are expressed as
\begin{align}
&\dot{\rho}_{de}+3H(1+w)\rho_{de}=-Q,\\
&\dot{\rho}_c+3H\rho_c=Q,\\
&\dot{\rho}_b+3H\rho_b=0,\\
&\dot{\rho}_r+4H\rho_r=0,
\end {align}
where the dot denotes a derivative with respect to time $t$, $H=\dot{a}/a$ is the Hubble parameter, $\rho_{\emph{i}}$ is the energy density of each component, for $i$ $=$ de, c, b, and r, respectively, and $Q$ is the energy transfer rate between DE and CDM. In this paper, we consider two interacting forms: $Q=\beta H\rho_c$ (denoted as $Q_1$) and $Q=\beta H\rho_{de}$ (denoted as $Q_2$), where $\beta$ is the dimensionless coupling.
For convenience, the I$\Lambda$CDM models with $Q_1$ and $Q_2$ are denoted as the I$\Lambda$CDM1 and I$\Lambda$CDM2 models, respectively, and the I$w$CDM models with $Q_1$ and $Q_2$ are denoted as the I$w$CDM1 and I$w$CDM2 models, respectively.

This paper is organized as follows. In Sec. \ref{methods}, we briefly review the two diagnostics, i.e., the statefinder hierarchy and the growth rate of structure. In Sec. \ref{results}, we explore the interacting dark energy models with the geometrical and structural diagnostics. Conclusion is given in Sec. \ref{concl}.

\section {Statefinder hierarchy and growth rate of structure}
\label{methods}

In this section, we will first review the statefinder hierarchy diagnostic and then describe the growth rate of structure in the interacting DE models. In particular, we present the ``parametrized post-Friedmann'' framework for interacting DE models, which is used to calculate the growth rate of structure in the I$\Lambda$CDM and I$w$CDM models.

\subsection{The statefinder hierarchy}
\label{method1}
The scale factor of the universe, $a(t)/a_0=(1+z)^{-1}$, can be Taylor expanded around the present epoch $t_0$ as follows:
\begin{equation}
\frac{a(t)}{a_0}=1+\sum\limits_{\emph{n}=1}^{\infty}\frac{A_{\emph{n}}(t_0)}{n!}[H_0(t-t_0)]^n,
\end {equation}
where
\begin{equation}
A_{\emph{n}}=\frac{a(t)^{(n)}}{a(t)H^n},~~n\in N,
\end {equation}
with $a(t)^{(n)}=d^na(t)/dt^n$. $A_{\emph{n}}$ ($n\geq3$) is transformed to statefinder hierarchy, $S^{(1)}_{\emph{n}}$, in order to obtain null diagnostic for the $\Lambda$CDM model, which is expressed as~\cite{Arabsalmani:2011fz}:
\begin{align}
&S^{(1)}_{3}=A_{3},\\
&S^{(1)}_{4}=A_{4}+3(1+q)\\
&S^{(1)}_{5}=A_{5}-2(4+3q)(1+q),~~\rm{etc.},
\end{align}
where the superscript $(1)$ just is a marked symbol, and $q$ is the deceleration parameter, $q=-A_2$.

In a spatially flat FRW universe consisting of only DE and non-relativistic matter, the statefinder hierarchy for $\Lambda$CDM is fixed to be $1$ during the cosmic expansion, namely, $S^{(1)}_{\emph{n}}|_{\Lambda \rm{CDM}}=1$. By using this diagnostic, one can distinguish easily the $\Lambda$CDM model from other DE models. However, in this paper, we consider a spatially flat FRW universe containing DE, CDM, baryons, and radiation. Obviously, in this case, $S^{(1)}_{\emph{n}}|_{\Lambda \rm{CDM}}$ is no longer a constant, but slowly changes with time during the cosmic expansion. Even so, the $\Lambda$CDM model is still treated as a reference model.

For the interacting DE scenario, we derive the expressions of $S^{(1)}_{3}$ and $S^{(1)}_{4}$ in terms of the model parameters:
\begin{equation}
S^{(1)}_{3}=1+\frac{9}{2}\Omega_{de}w(1+w)-\frac{3}{2}w'\Omega_{de}+2\Omega_r+\frac{3wQ}{2H\rho},
\end{equation}
\begin{equation}
\begin{aligned}
S^{(1)}_{4}=&1-\frac{9}{4}w\Omega_{de}^2[3w(1+w)-w']-\frac{3}{4}\Omega_{de}[w(21+39w+18w^2)+\Omega_r(7w+3w^2-w')\\
&-(13+18w)w'+2w'']-\Omega_r(9+\Omega_r)-\frac{3wQ}{2H\rho}(2+3w)+\frac{3wQ'}{2H\rho}+\frac{3w'Q}{H\rho}.
\end{aligned}
\end{equation}

where the prime denotes the derivative with respect to $x=\ln a$ and $\rho=\sum \rho_i$. The above expressions are applicable for all the interacting DE models with arbitrary $w(z)$ and $Q$. For the I$w$CDM models, one should let $w={\rm const}$ in the above formulae. For the I$\Lambda$CDM models, one substitutes $w=-1$ in the above formulae.

\subsection{The growth rate of structure}
\label{method2}
The fractional growth parameter $\epsilon (z)$~\cite{Acquaviva:2008qp,Acquaviva:2010vr} is defined as
\begin{equation}
\epsilon(z)=\frac{f(z)}{f_{\Lambda CDM}(z)},
\end {equation}
where $f(z)=d\ln\delta/d\ln a$ is the growth rate of structure. Here, $\delta=\delta \rho_m/\rho_m$, with $\rho_m$ and $\delta\rho_m$ being the energy density and the density perturbation of matter (including CDM and baryons), respectively.

If the matter density perturbation is linear and there is no interaction between DE and CDM, the perturbation equation at late times is written as:
\begin{equation}\label{DEpert}
\ddot{\delta}+2H\dot{\delta}=4\pi G\rho_m\delta ,
\end {equation}
where $G$ is the Newton's gravitational constant.
Consequently, the growth rate of the linear density perturbation is approximatively given by~\cite{Wang:1998gt}:
\begin{equation}
f(z)\simeq \Omega_m(z)^\gamma,\label{eq-fz}
\end {equation}
\begin{equation}
\gamma(z)=\frac{3}{5-\frac{w}{1-w}}+\frac{3(1-w)(1-\frac{3}{2}w)}{125(1-\frac{6}{5}w)^3}(1-\Omega_m(z)),\label{eq-r}
\end {equation}
where $\Omega_{m}(z)\equiv \rho_{m}(z)/{3M^2_p}H(z)^2$ is the fractional density of matter, $w$ is either a constant or varies slowly with time. For the $\Lambda$CDM model, $\gamma\simeq0.55$ and $\epsilon(z)=1$~\cite{Wang:1998gt,Linder:2005in}. For other models, the values of $\epsilon(z)$ exhibit differences from $\Lambda$CDM, which is the reason why the fractional growth parameter $\epsilon (z)$ is used as a diagnostic.

However, for an interacting DE model, the growth rate $f(z)$ cannot be simply parameterized by Eqs. (\ref{eq-fz}) and (\ref{eq-r}) \cite{CalderaCabral:2009ja}. For more details of calculating the growth rate in interacting DE models, we refer the reader to \cite{CalderaCabral:2009ja}. But in this paper we calculate the growth rate within the ``parametrized post-Friedmann'' (PPF) framework for interacting dark energy \cite{Li:2014eha,Li:2014cee}. One of the advantages of the PPF approach is that it could resolve the problem of early-time super-horizon-scale perturbation divergence in the interacting DE models. We will briefly present the PPF framework for interacting DE in the following.

In an interacting DE model, the conservation laws become
\begin{equation}
\label{eqn:energyexchange} \nabla_\nu T^{\mu\nu}_{\emph{I}} = Q^\mu_{\emph{I}}\,, \quad\quad
 \sum_{\emph{I}} Q^\mu_{\emph{I}} = 0,
\end{equation}
where $Q^\mu_{\emph{I}}$ denotes the energy-momentum transfer of $I$ fluid. Generally, $Q^\mu_{\emph{I}}$ can be split as 
\begin{equation}
Q_{\mu}^I  = a\big( -Q_{\emph{I}}(1+AY) - \delta Q_{\emph{I}}Y,\,[ f_{\emph{I}}+ Q_{\emph{I}} (v-B)]Y_{\emph{i}}\big),\label{eq:Qenergy}
\end{equation}
where $\delta Q_{\emph{I}}$ and $f_{\emph{I}}$ are the energy transfer perturbation and momentum transfer potential of $I$ fluid, respectively, $A$ and $B$ are functions describing the perturbed metric, $v$ denotes the velocity perturbation of total matters, and $Y$ and $Y_{\emph{i}}$ are the eigenfunctions of the Laplace operator and its covariant derivatives, respectively. Then, Eqs.~(\ref{eqn:energyexchange}) and (\ref{eq:Qenergy}) give the following two conservation equations,
\begin{align}
\delta\rho'_{\emph{I}}+3(\delta\rho_{\emph{I}}+\delta p_{\emph{I}})+(\rho_{\emph{I}}+p_{\emph{I}})(k_{\emph{H}} v_{\emph{I}}+3H'_{\emph{L}})
 =\frac{1}{H}(\delta Q_{\emph{I}}-AQ_{\emph{I}}),\label{15} \\
\frac{[a^4(\rho_{\emph{I}}+p_{\emph{I}})(v_{\emph{I}}-B)]'}{a^4k_{\emph{H}}}-\delta p_{\emph{I}}+\frac{2}{3}p_{\emph{I}}\Pi_{\emph{I}}-(\rho_{\emph{I}}+p_{\emph{I}})A
 =\frac{a}{k}[Q_{\emph{I}}(v-B)+f_{\emph{I}}],\label{16}
 \end{align}
where $k_{\emph{H}}=k/(Ha)$ with $k$ the wave number, the prime denotes the derivative with respect to $x=\ln a$, $H_{\emph{L}}$ also denotes metric perturbation, and $\delta p_{\emph{I}}$ and $\Pi_{\emph{I}}$ are pressure perturbation and anisotropic stress of $\emph{I}$ fluid, respectively. The values of $\delta Q_I$ and $f_I$ depend on the covariant form of $Q^\mu_{\emph{I}}$. In our work, we choose $Q^{\mu}_c  = -Q^{\mu}_{de}=Q u^{\mu}_c$ with $u^{\mu}_c= a^{-1}\big(1-AY,\,v_cY^i \big)$, so that there is no momentum transfer in the CDM frame. Comparing with Eq.~(\ref{eq:Qenergy}), we can obtain $\delta Q_{de}=-\delta Q_{c}=-\beta H\rho_c\delta_c$ for $Q=\beta H\rho_c$, $\delta Q_{de}=-\delta Q_{c}=-\beta H\rho_{de}\delta_{de}$ for $Q=\beta H\rho_{de}$, and $f_{de}=-f_c=\beta H\rho_c(v-v_c)$.

In above equations, $\Pi_i$ generally vanish for CDM and DE, and the metric perturbations are decided by a specific gauge and the Einstein equations,
 \begin{align}
& {H_{\emph{L}}}+ {{H_{\emph{T}}} \over 3} +   {B \over k_{\emph{H}}}-  {H_{\emph{T}}' \over k_{\emph{H}}^2}= { 4\pi Ga^2 \over k^2}   \left[ {\delta \rho} + 3  (\rho+p){{v}-
{B} \over k_{\emph{H}} }\right] \,,\label{eqn:einstein1}\\
& {A} - {H_{\emph{L}}'}
- { H_{\emph{T}}' \over 3} - {K \over (aH)^2} \left( {B \over k_{\emph{H}}} - {H_{\emph{T}}' \over k_{\emph{H}}^2} \right)=  {4\pi G \over H^2 } (\rho+p){{v}-{B} \over k_{\emph{H}}} \,.
\label{eqn:einstein2}
\end{align}
To complete the perturbation systems of CDM and DE, we also need an extra condition on $\delta p_{\emph{I}}$. In uncoupled DE case, one often treats DE as a nonadiabatic fluid and calculates $\delta p_{de}$ in terms of its rest-frame sound speed, which, however, induces the large-scale instability for the interacting DE scenario \cite{Valiviita:2008iv,He:2008si}. To avoid this large-scale instability, we here handle the perturbation evolutions using the PPF framework for interacting DE model established in \cite{Li:2014eha,Li:2014cee}.

Instead of calculating $\delta p_{de}$ in terms of rest-frame sound speed of DE, the PPF approach completes the perturbation system by establishing a direct relationship between $V_{de} - V_T$ and $V_T$ on the large scales, where $V_{de}$ and $V_T$ are the velocity perturbation of DE and total matters except DE in the comoving gauge. This relationship can be parametrized by a function $f_\zeta(a)$ as \cite{Hu:2008zd,Fang:2008sn}
\begin{equation}
\lim_{k_{\emph{H}} \ll 1}
 {4\pi G \over H^2} (\rho_{de} + p_{de}) {V_{de} - V_T \over k_{\emph{H}}}
= - {1 \over 3}  f_\zeta(a) k_{\emph{H}} V_T.\label{eq:DEcondition}
\end{equation}
With the help of this extra condition in combination with Eqs. (\ref{15})--(\ref{eqn:einstein2}), we can finally solve all the perturbation equations and numerically get the growth rate $f(z)$ in the interacting DE models. For more details, see \cite{Li:2014eha,Li:2014cee}.

Besides $\epsilon(z)$, we can use a composite null diagnostic (CND), $\{S^{(1)}_{\emph{n}},\epsilon\}$, combining $\epsilon(z)$ with statefinder hierarchy $S^{(1)}_{\emph{n}}$~\cite{Arabsalmani:2011fz}.

Before we apply the CND to interacting DE models, we wish to make some discussions on the consistency relation between expansion history and growth of structure for interacting DE. As is well known, for the non-interacting DE, the matter perturbations are entirely determined by the expansion background, as described by Eq. (\ref{DEpert}). This provides a general strategy to distinguish between DE and modified gravity (MG) scenarios: a mismatch (distinct tension) in DE parameter space for constraints from geometrical measurements (e.g., luminosity distance and angular diameter distance) and growth-of-structure measurements (e.g., linear growth rate) points to a modification of GR. In general, for MG, the growth of structure is usually scale dependent, which is a feature not seen in smooth DE models.\footnote{For the MG scenario, in the quasi-static Newtonian regime, we have $\ddot{\delta}+2H\dot{\delta}-{4\pi\tilde{G}\over(\Phi/\Psi)}\rho_m\delta=0$. From this equation one sees how the combination of $\tilde{G}$ and $\Phi/\Psi$ alters the linear matter perturbation. If these parameters have a scale dependence, then the linear growth function becomes scale dependent.} Now, a question arises: Is the growth of structure in the interacting DE scenario also scale dependent? Actually, this issue has been addressed in \cite{CalderaCabral:2009ja} and the answer is ``NO''.

Here we quote the results of \cite{CalderaCabral:2009ja}. For the $Q^\mu=\beta H\rho_c u_c^\mu$ model, the second-order evolution equation for $\delta_c$ is
\begin{equation}\label{IDEpert1}
\ddot{\delta}_c+2H\dot{\delta}_c-4\pi G(\rho_c\delta_c+\rho_b\delta_b)=0.
\end{equation}
Though this equation has the same form as that for the non-interacting case, the solutions $\delta_c$ are different because the background terms $H$ and $\rho_c$ evolve differently. For the $Q^\mu=\beta H\rho_{de} u_c^\mu$ model, the evolution equation of $\delta_c$ gets two apparent modifications: a modified Hubble friction term and a modified source term, expressed as a modified effective Newton constant for CDM:
\begin{equation}\label{IDEpert2}
\ddot{\delta}_c+2H\left(1+\beta{\rho_{de}\over \rho_c}\right)\dot{\rho}_c-4\pi G_{eff}\rho_c\delta_c-4\pi G\rho_b\delta_b=0,
\end{equation}
where
\begin{equation}\label{Geff}
{G_{eff}\over G}=1-{2\beta\over 3a}{\rho_{de}\over\rho_c}\left\{2-3w-\beta\left(1+{\rho_{de}\over \rho_c}\right)\right\}.
\end{equation}
We see clearly from Eqs. (\ref{IDEpert1})--(\ref{Geff}) that, for these interacting DE models, the growth of structure is still scale-independent. Also, from these equations, we find that though the direct non-gravitational interaction between DE and CDM is introduced, the matter perturbations could still be derived from background observables, such as luminosity distance and angular diameter distance data.

In \cite{Alam:2008at}, Alam, Sahni and Starobinsky proposed to reconstruct cosmological matter perturbations using observed luminosity distance and angular diameter distance data. They considered non-interacting DE cases and thus used Eq. (\ref{DEpert}) to perform this reconstruction. Here we have shown that, if one wishes to perform such a reconstruction for the interacting DE cases (I$w$CDM1 and I$w$CDM2), one should use Eqs. (\ref{IDEpert1})--(\ref{Geff}) (note that the evolution equation for $\delta_b$ is not changed and thus is not given here). If the results of these reconstructions for both non-interacting DE and interacting DE cases are in tension with the actual observational data of growth rate of structure from, e.g., redshift space distortions (RSD), then this would give a hint for the MG scenario.

So far, we have clarified that in the interacting DE models the growth of structure is still consistent with the expansion history. Hence, in these cases, both the statefinder hierarchy and fractional growth parameter $\epsilon$ could be inferred from the same observational data, e.g., supernovae and BAO observations.

\section{Exploration of interacting dark energy with geometrical and structural diagnostics}
\label{results}

In this section, we explore the deviations from $\Lambda$CDM in the interacting DE models by using the diagnostics of statefinder hierarchy $S_n^{(1)}$ and fractional growth parameter $\epsilon (z)$. Since we only care about the impacts of the EoS parameter $w$ and the coupling parameter $\beta$, we fix the other model parameters. For all models, the present-day fractional density parameters of CDM, baryons, and radiation are fixed to be: $\Omega_{c0}=0.23$, $\Omega_{b0}=0.04$, and $\Omega_{r0}=2.469\times10^{-5}h^{-2}(1+0.2271N_{eff})$ (with the Hubble constant $h=0.7$ and the effective number of neutrino species $N_{eff}=3.046$). For properly choosing the values of EoS and coupling, we refer to the literature~\cite{Geng:2014hoa,Geng:2015ara,Li:2013bya,Wang:2014oga}. We will first test the impact of $w$ in the I$w$CDM models. In this test, we fix the coupling to be $\beta=0.02$ and consider the cases of $w=-0.9$, $-1$, and $-1.1$. Then, we wish to test the impact of $\beta$ in the I$\Lambda$CDM models. In this test (with $w=-1$), we consider the cases of $\beta=-0.04$, $-0.02$, 0, 0.02, and 0.04.

\subsection{Testing the impacts of $w$ and $\beta$ in the I$w$CDM models}

In this subsection, we will have a look at the impacts of $w$ and $\beta$ in the I$w$CDM models using the diagnostics of statefinder hierarchy and fractional growth parameter.

\begin{figure*}[htbp]
\centering
\includegraphics[scale=0.3]{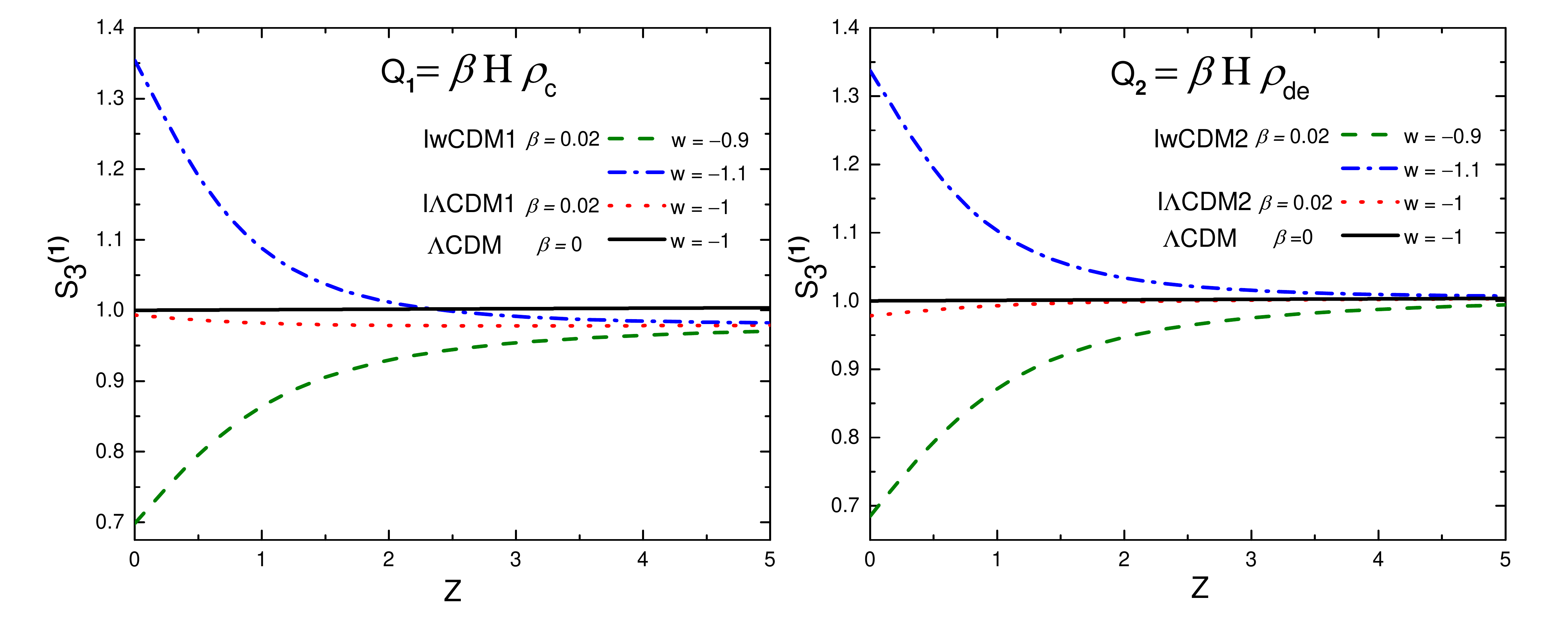}
\caption{\label{fig1} Evolutions of $S^{(1)}_3$ versus redshift $z$ for the I$w$CDM model with fixed coupling $\beta=0.02$. The $S^{(1)}_3(z)$ curve of the $\Lambda$CDM model is also shown for comparison. The left panel is for the model with $Q_1=\beta H\rho_c$ and the right panel is for the model with $Q_2=\beta H\rho_{de}$.}
\end{figure*}

\begin{figure*}[htbp]
\centering
\includegraphics[scale=0.3]{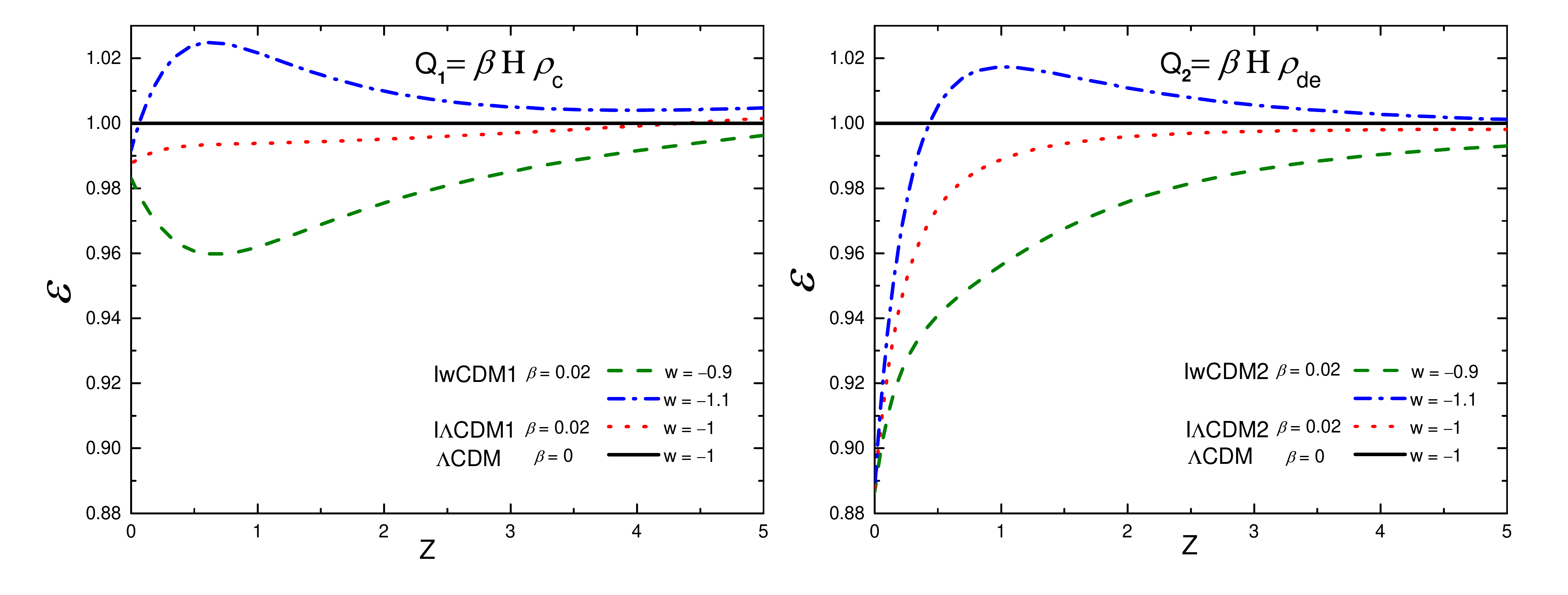}
\caption{\label{fig2}  Evolutions of $\epsilon$ versus redshift $z$ for the I$w$CDM model with fixed coupling $\beta=0.02$. The $\epsilon(z)$ curve of the $\Lambda$CDM model is also shown for comparison. The left panel is for the model with $Q_1=\beta H\rho_c$ and the right panel is for the model with $Q_2=\beta H\rho_{de}$.}
\end{figure*}

First, let us diagnose the I$w$CDM models with $Q_1$ and $Q_2$ in the light of the statefinder parameter $S_3^{(1)}$. In Fig. \ref{fig1}, we plot the evolutions of $S^{(1)}_3$ versus redshift $z$ for the two I$w$CDM models with fixed coupling $\beta=0.02$ and varying EoS values of $w=-0.9$, $-1$, and $-1.1$. Note that I$\Lambda$CDM is a specific case for I$w$CDM with $w=-1$, as labeled in Fig. \ref{fig1}. Since we wish to explore the deviations from the reference model, $\Lambda$CDM, we also show the $\Lambda$CDM model as a specific case with $\beta=0$ and $w=-1$ in this figure for a direct comparison. We find that using only the $S^{(1)}_3$ can easily distinguish the different cases of I$w$CDM with fixed $\beta$ and varying $w$. For both I$w$CDM1 and I$w$CDM2, we find that the departures between the three lines (with $\beta=0.02$) are prominent in the low redshifts up to about $z=3$. But when we compare the two lines with the same $w$ (i.e., $w=-1$) and different $\beta$ (i.e., $\beta=0$ and 0.02), we find that they are nearly degenerate in the whole redshift range. That is to say, it is difficult to probe the deviation of I$\Lambda$CDM from $\Lambda$CDM using only the statefinder diagnostic $S^{(1)}_3$. We will use the higher hierarchy of statefinder, $S^{(1)}_4$, to further probe the deviation from $\Lambda$CDM in the next subsection.

Next, we diagnose the I$w$CDM models in the light of the fractional growth parameter $\epsilon$. In Fig. \ref{fig2}, we plot the evolutions of $\epsilon$ versus $z$ for the two I$w$CDM models where the parameter values are taken to be the same with those in Fig. \ref{fig1}. We find that using only the structure growth diagnostic, it is also easy to distinguish the cases of I$w$CDM with fixed $\beta$ and varying $w$. We find that for both $Q_1$ and $Q_2$ models the $\epsilon$ values in the redshift $z=0$ are nearly degenerate, but they are distinguished well in higher redshifts, with the best distinguishing window of $z\simeq 0.5$--1.5. Furthermore, we compare the I$\Lambda$CDM and the $\Lambda$CDM cases, i.e., the lines with $\beta=0.02$ and 0 in this figure. We find that the difference of them is bigger in the lower redshifts, and the I$\Lambda$CDM2 case (right panel) is better than the I$\Lambda$CDM1 case (left panel).

In both Figs.~\ref{fig1} and~\ref{fig2}, we find that there exist degeneracies in the high-redshift region, no matter which diagnostic is used. But this does not influence our diagnosis, since the observational data are mainly within the low-redshift region. So we pay more attention to distinguish models in the low-redshift region. The current values of $S^{(1)}_{\emph{n}}$ and $\epsilon$, i.e., $S^{(1)}_{\emph{n}to}$ and $\epsilon _0$, can be viewed as discriminators for testing various cosmological models. For the different interacting DE models, the values of $S^{(1)}_{3to}$, $S^{(1)}_{4to}$, and $\epsilon _0$ are listed in Table~\ref{tab1}, which can supply assistant information.

From the above analysis, we find that the impact of $w$ is much stronger than that of $\beta$ in the I$w$CDM models in both geometrical and structure growth diagnostics. Actually, we are more interested in the one-parameter extension to $\Lambda$CDM, i.e., the I$\Lambda$CDM scenario. Thus, in the next subsection, we will probe the deviation of I$\Lambda$CDM from $\Lambda$CDM regarding the extra parameter $\beta$, by using the combination of geometrical and structural diagnostics.

\begin{table*}[tbp]
\centering
\begin{tabular}{cccccccccccccc}
\hline
EoS&&$w$&&\multicolumn{5}{c}{$w=-1$}&&\multicolumn{1}{c}{$w=-0.9$}&&\multicolumn{1}{c}{$w=-1.1$}&\\
\cline{5-9}\cline{11-11}\cline{13-13}
coupling&&$\beta$&&$0.04$&$0.02$&$0$&$-0.02$&$-0.04$&&$0.02$&&$0.02$&\\
\hline
&&$S^{(1)}_{3to}$&&$0.986$&$0.993$&$1.000$&$1.007$&$1.014$&&$0.698$&&$1.354$&\\
$Q_1$&&$S^{(1)}_{4to}$&&$1.032$&$1.016$&$0.999$&$0.983$&$0.966$&&&&&\\
&&$\epsilon _{0}$&&$0.983$&$0.988$&$1$&$0.998$&$1.003$&&$0.983$&&$0.992$&\\
\hline
&&$S^{(1)}_{3to}$&&$0.956$&$0.978$&$1.000$&$1.022$&$1.044$&&$0.685$&&$1.337$&\\
$Q_2$&&$S^{(1)}_{4to}$&&$0.975$&$0.987$&$0.999$&$1.013$&$1.027$&&&&&\\
&&$\epsilon _{0}$&&$0.78$2&$0.887$&$1$&1.098&1.204&&$0.886$&&$0.887$&\\
\hline
\end{tabular}
\caption{\label{tab1} The current values of the statefinder hierarchy and the fractional growth parameter, $S^{(1)}_{3to}$, $S^{(1)}_{4to}$, and $\epsilon _0$, for the interacting DE models.}
\end{table*}

\subsection{Probing deviations from $\Lambda$CDM in the I$\Lambda$CDM models}

The I$\Lambda$CDM is a one-parameter extension to the $\Lambda$CDM, with the extra parameter $\beta$. We are indeed interested in probing the deviation from $\Lambda$CDM in terms of the coupling $\beta$. So in this subsection we will have a closer look at the I$\Lambda$CDM models with the statefinder hierarchy and the growth rate of structure.

\begin{figure*}[htbp]
\centering
\includegraphics[scale=0.6]{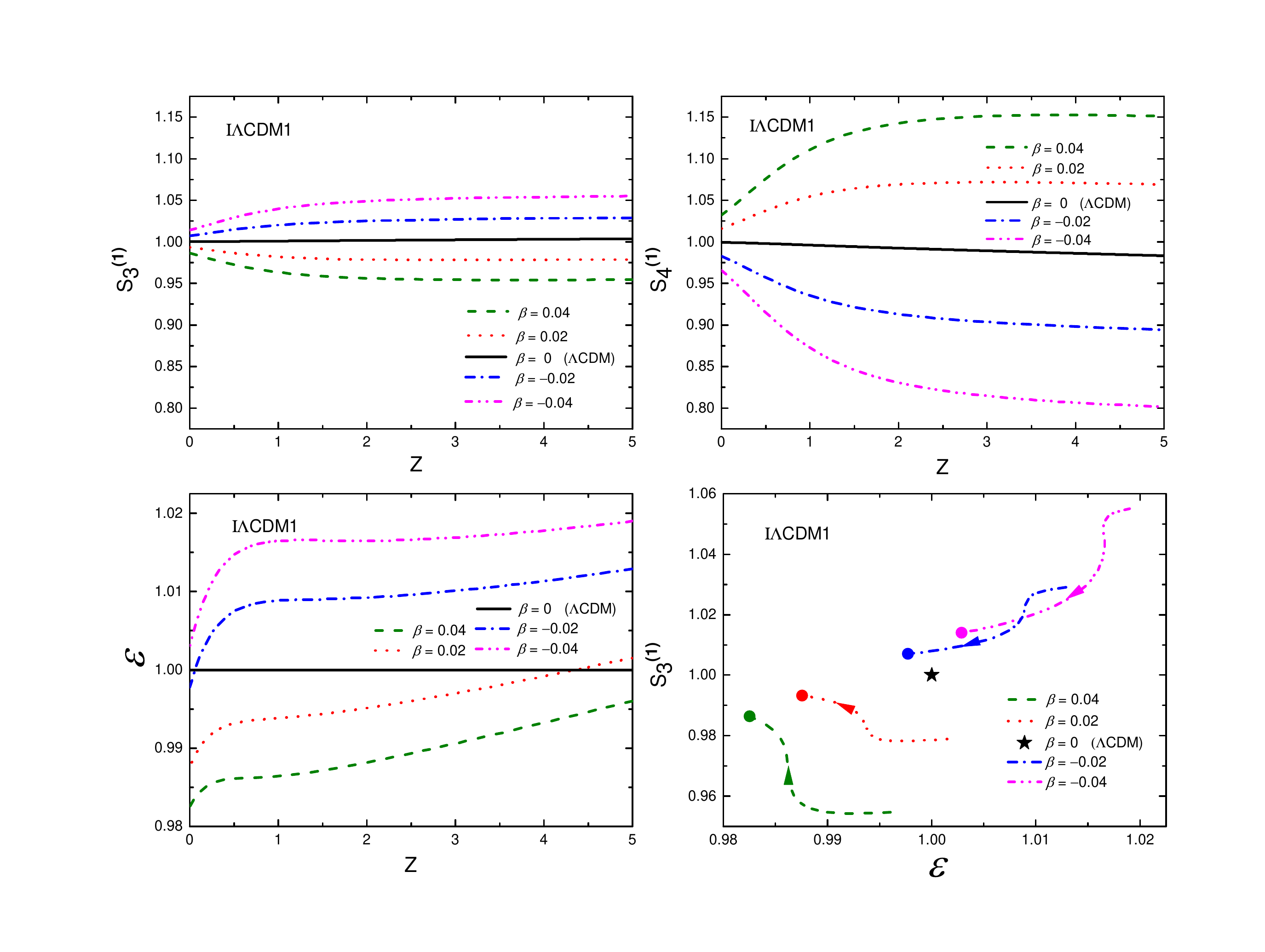}
\caption{\label{fig3} Evolution curves of $S^{(1)}_3(z)$, $S^{(1)}_4(z)$, and $\epsilon(z)$, and evolutionary trajectories of $S^{(1)}_3(\epsilon)$ for the I$\Lambda$CDM1 model ($Q_1=\beta H\rho_c$) with $\beta=0$  ($\Lambda$CDM), $\pm0.02$, and $\pm0.04$. In the $S^{(1)}_3$--$\epsilon$ plane, the current values of $\{S^{(1)}_3,\epsilon\}$ of the I$\Lambda$CDM1 model are marked by the round dots. The $S^{(1)}_3(\epsilon)$ trajectory of the $\Lambda$CDM model is approximately taken to be a point $\{1,1\}$, shown as a star. The arrows indicate the evolution directions of the $S^{(1)}_3(\epsilon)$ trajectories.}
\end{figure*}

\begin{figure*}[htbp]
\centering
\includegraphics[scale=0.6]{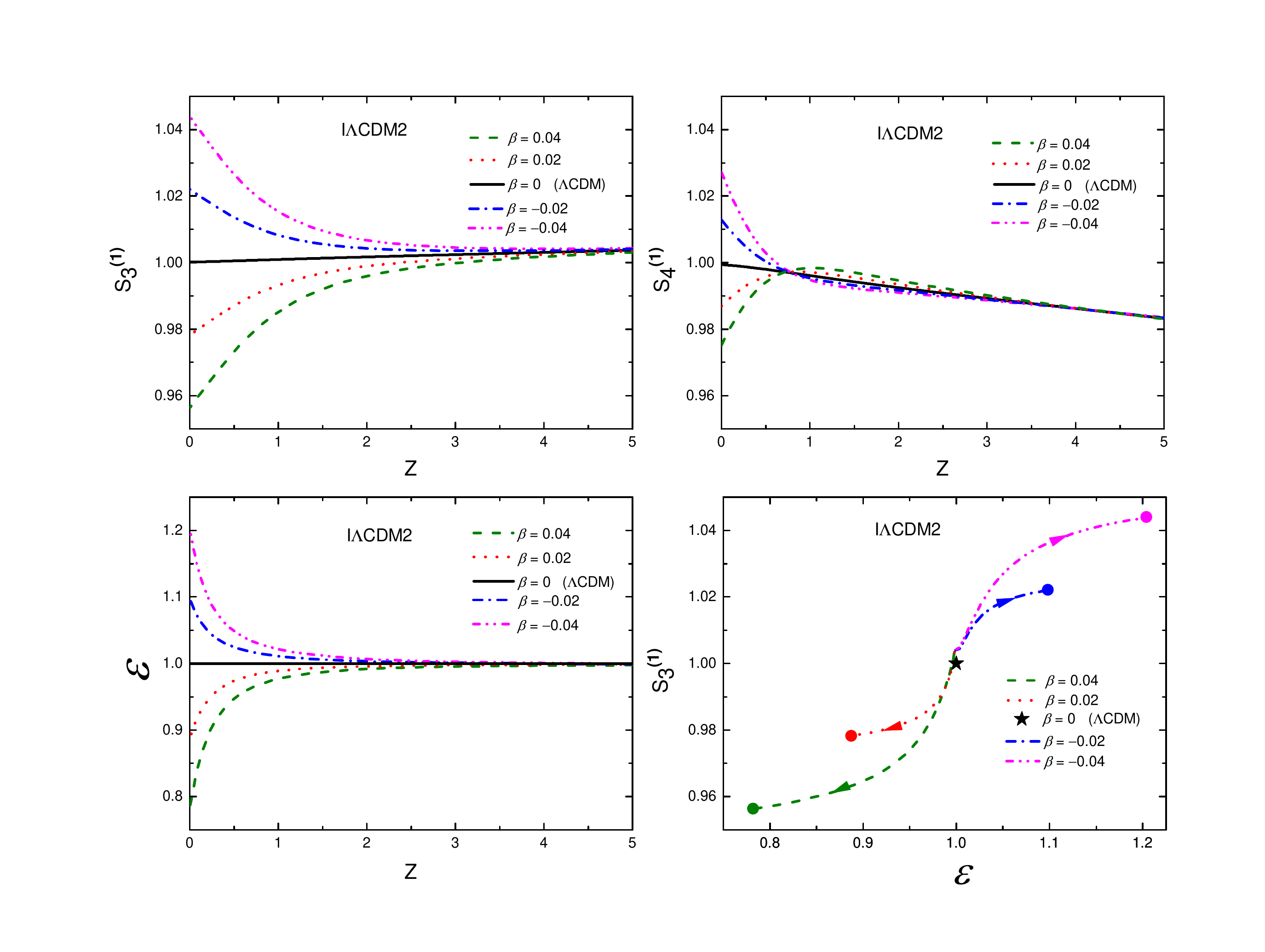}
\caption{\label{fig4} Evolution curves of $S^{(1)}_3(z)$, $S^{(1)}_4(z)$, and $\epsilon(z)$, and evolutionary trajectories of $S^{(1)}_3(\epsilon)$ for the I$\Lambda$CDM2 model ($Q_2=\beta H\rho_{de}$) with $\beta=0$  ($\Lambda$CDM), $\pm0.02$, and $\pm0.04$. In the $S^{(1)}_3$--$\epsilon$ plane, the current values of $\{S^{(1)}_3,\epsilon\}$ of the I$\Lambda$CDM2 model are marked by the round dots. The $S^{(1)}_3(\epsilon)$ trajectory of the $\Lambda$CDM model is approximately taken to be a point $\{1,1\}$, shown as a star. The arrows indicate the evolution directions of the $S^{(1)}_3(\epsilon)$ trajectories. }
\end{figure*}

In this case, the deviation from $\Lambda$CDM only comes from the impact of the coupling $\beta$. To perform a diagnostic analysis, we take $\beta=0$ ($\Lambda$CDM), $\pm 0.02$, and $\pm 0.04$, for the I$\Lambda$CDM scenario. For the statefinder hierarchy diagnostic, we consider the evolutions of both $S^{(1)}_3(z)$ and $S^{(1)}_4(z)$. For the structure growth diagnostic, we consider the evolution of $\epsilon(z)$. Moreover, we also perform a CND $\{S^{(1)}_3,\epsilon\}$ for the models.

We first discuss the I$\Lambda$CDM1 model (with $Q_1=\beta H\rho_c$). In Fig.~\ref{fig3}, we plot the evolution curves of $S^{(1)}_3(z)$, $S^{(1)}_4(z)$, and $\epsilon(z)$, as well as the evolutionary trajectories in the $S^{(1)}_3$--$\epsilon$ plane, for the model. In the $S^{(1)}_3$--$\epsilon$ plane, the current values of $\{S^{(1)}_3, \epsilon\}$ of the I$\Lambda$CDM1 model are marked by the round dots, and the arrows indicate the evolution directions of various cases for the model. It should be noticed that $S^{(1)}_3\simeq1$ and $\epsilon=1$ for the $\Lambda$CDM model during the whole evolution history. As a result, the trajectory of $\Lambda$CDM is a tiny line segment in the $S^{(1)}_3$--$\epsilon$ plane. For simplification, we replace the short $S^{(1)}_3(\epsilon)$ trajectory of $\Lambda$CDM approximately with a point $\{1,1\}$ shown as a star. We find that, for the curves of $S^{(1)}_3(z)$, the separations between them are rather near, indicating a near degeneration. But for the curves of $S^{(1)}_4(z)$, it is easy to find that the separations between them become much farther than in the $S^{(1)}_3(z)$ case. Specifically, the maximum deviations from $\Lambda$CDM for $S^{(1)}_3(z)$ and $S^{(1)}_4(z)$ are about 5\% and 15\%, respectively, showing that $S^{(1)}_4(z)$ could well measure the deviation from $\Lambda$CDM and differentiate the geometrical degeneracies in the model. In the $\epsilon(z)$ panel, we find that the maximum deviation from $\Lambda$CDM for $\epsilon(z)$ is only about 2\%. Thus the $\epsilon$ diagnostic only provides tiny differentiation. In the $S^{(1)}_3$--$\epsilon$ plane, the departures from $\Lambda$CDM, from the points of view of both expansion history and growth of structure, could be directly measured in this plane. The $S^{(1)}_3(\epsilon)$ trajectories also exhibit distinctive features for the various cases of the I$\Lambda$CDM1 model.

Next, we discuss the I$\Lambda$CDM2 model (with $Q_2=\beta H\rho_c$). The evolution curves of $S^{(1)}_3(z)$, $S^{(1)}_4(z)$, and $\epsilon(z)$, and the evolutionary trajectories in the $S^{(1)}_3$--$\epsilon$ plane for the I$\Lambda$CDM2 model, are plotted in Fig. \ref{fig4}. When we look at the $S^{(1)}_3(z)$ curves, we find that they are highly degenerate in the high redshift region, but they can separate from each other in the low redshift region. Even though in the low redshift range the $S^{(1)}_3(z)$ curves can be discriminated to some extent, the maximum deviations from $\Lambda$CDM for $S^{(1)}_3(z)$ and $S^{(1)}_4(z)$ are only about 4\% and 3\%, respectively. Comparing to the I$\Lambda$CDM1 case (notice the different scales in Figs. \ref{fig3} and \ref{fig4} and refer also to the concrete values in Table \ref{tab1}), we find that actually this statefinder diagnostic only provides a mild differentiation and the curves are degenerate in some degree. Interestingly, we find that in this model the $S^{(1)}_4(z)$ curves exhibit even stronger degeneracy than the case of $S^{(1)}_3(z)$, as shown in Fig. \ref{fig4}. Thus, for the I$\Lambda$CDM2 model, the $S^{(1)}_4$ diagnostic does not provide any useful help for differentiating the geometrical degeneracies. However, we find that for this model the $\epsilon(z)$ diagnostic could well discriminate the various cases of the model and finely measure the deviations from $S^{(1)}_3$, in the low redshift region (the maximum deviation from $\Lambda$CDM for $\epsilon(z)$ is about 20\%). In the high redshift region, the I$\Lambda$CDM2 model is highly degenerate with $\Lambda$CDM. Therefore, for the I$\Lambda$CDM2 model, combining the diagnostics of expansion history and growth of structure becomes extremely important. We thus perform a CND $\{S^{(1)}_3,\epsilon\}$ for the model. We find that in the $S^{(1)}_3$--$\epsilon$ plane the trajectories exhibit distinctive features and the departures from $\Lambda$CDM could be well measured. The power of CND is clearly shown in this case.

In the above discussions, the deviation from $\Lambda$CDM for each diagnostic parameter has been measured, theoretically. Now we wish to compare these deviations with the uncertainties on diagnostic parameters given by observations. To do this, we should first estimate the parameter space of the diagnostic parameters using observational data. However, this kind of work needs a series of calculations that are not the main focus of this paper. So here we quote the observed uncertainties on diagnostic parameters obtained in previous works. Uncertainties on $S^{(1)}_3$ and $S^{(1)}_4$ from current observations or estimated from simulations can be found in~\cite{Alam:2003sc,Ishida:2006vk,Rapetti:2006fv,Wang:2009gv,Xu:2010hq,Capozziello:2011tj,Capozziello:2013wha}. It was found that they are rather difficult to measure---for most of the fit results, the uncertainties on $S^{(1)}_3$ and $S^{(1)}_4$ at $z=0$ are not less than 6\% and 20\%, respectively. These uncertainties on $S^{(1)}_3$ and $S^{(1)}_4$ are larger than the maximum differences [5\% (4\%) and 15\% (3\%)] between the $\Lambda$CDM model and the I$\Lambda$CDM1 (I$\Lambda$CDM2) model, indicating that neither the I$\Lambda$CDM1 model nor the I$\Lambda$CDM2 model can be well distinguished from the $\Lambda$CDM model. On the other hand, the fit results of $\epsilon$ can be found in~\cite{Alam:2008at,Acquaviva:2008qp,Acquaviva:2010vr}, where the uncertainty on $\epsilon$ can be constrained to about 5\%. With this precision, the I$\Lambda$CDM2 model, where the deviation from the $\Lambda$CDM model for $\epsilon$ is about 20\%, can be well distinguished. However, the I$\Lambda$CDM1 model still cannot be distinguished (the deviation for $\epsilon$ in this model is about 2\%). In summary, the diagnostic parameters $S^{(1)}_n$ and $\epsilon$ still do not have sufficient precisions to completely distinguish I$\Lambda$CDM models from the $\Lambda$CDM model, currently. Note that their precisions may be further degraded, if one simultaneously constrains $S^{(1)}_n$ and $\epsilon$ using the same observations, due to the correlations between $S^{(1)}_n$ and $\epsilon$. This situation somehow weakens the power of CND $\{S^{(1)}_3,\epsilon\}$ considered in this paper. Nevertheless, the discussion of this paper is still meaningful, as it points out what precisions of the diagnostic parameters should be achieved to distinguish the I$\Lambda$CDM models from the $\Lambda$CDM model. We expect that future observations could measure the diagnostic parameters with such precisions.

\section {Conclusion}
\label{concl}

Currently, the $\Lambda$CDM cosmology provides an excellent description of various cosmological/astrophysical observations. Thus, any new physics beyond the $\Lambda$CDM cosmology, if detected and confirmed, would be viewed as a major breakthrough in cosmology and fundamental physics. The simplest extensions to the base $\Lambda$CDM cosmology regarding DE component includes $w$CDM and I$\Lambda$CDM. The former considers the model in which the cosmological constant $\Lambda$ is replaced by some DE field/fluid with constant $w$, and the latter considers the model in which the vacuum energy ($\Lambda$) directly couples to CDM in some physically profound way, and both are one-parameter extension to $\Lambda$CDM. The mixture of the two is called the I$w$CDM model.

In this paper, our focus is the I$\Lambda$CDM cosmology. We wish to explore the deviation of I$\Lambda$CDM from $\Lambda$CDM regarding the extra coupling parameter $\beta$ by using the diagnostics of both statefinder hierarchy and growth rate of structure. But, to make our discussion more generic, we begin with the more general model, namely, the I$w$CDM scenario. We consider two interacting form in this paper: $Q_1=\beta H\rho_c$ and $Q_2=\beta H\rho_{de}$. For the statefinder hierarchy, we derived the analytical expressions of $S^{(1)}_{3}$ and $S^{(1)}_{4}$ in terms of cosmological parameters for interacting dark energy. For the growth rate of structure, we employed the PPF theoretical framework for interacting dark energy to numerically obtain $\epsilon(z)$ for the models considered in this paper. We tested the I$w$CDM models with $Q_1$ and $Q_2$ using the $S^{(1)}_{3}$ and $\epsilon$ diagnostics. We found that in both the geometrical and structure growth diagnostics the impact of $w$ is much stronger than that of $\beta$ in the I$w$CDM models.

We wish to have a closer look at the I$\Lambda$CDM scenario that is a one-parameter extension to $\Lambda$CDM with the statefinder hierarchy and the growth rate of structure. We plotted the evolution curves of $S^{(1)}_3(z)$, $S^{(1)}_4(z)$, and $\epsilon(z)$, as well as the evolutionary trajectories in the $S^{(1)}_3$--$\epsilon$ plane, for the two I$\Lambda$CDM models with $Q_1$ and $Q_2$. We found that, for the model with $Q_1$, the geometrical diagnostic $S^{(1)}_4$ could well measure the deviation from $\Lambda$CDM and finely differentiate degeneracies in the model. But for the model with $Q_2$, we found that neither $S^{(1)}_3$ nor $S^{(1)}_4$ diagnostic could provide useful help in discriminating various degenerate cases in the model, while the $\epsilon$ diagnostic is fairly helpful for the model. We found that for both models the evolutionary trajectories in the $S^{(1)}_3$--$\epsilon$ plane exhibit distinctive features and the departures from $\Lambda$CDM could be well measured. Thus we have shown that the composite null diagnostic $\{S^{(1)}_3,\epsilon\}$ is a promising tool for diagnosing the I$\Lambda$CDM models. Besides, we also compare our results with the observed uncertainties on diagnostic parameters. We found that the diagnostic parameters still do not have sufficient precisions to completely distinguish I$\Lambda$CDM models from the $\Lambda$CDM model, currently. Anyway, our work gives what precisions of measurements should be achieved to distinguish the I$\Lambda$CDM models from the $\Lambda$CDM model.

\acknowledgments

This work was supported by the National Natural Science Foundation of China under Grant No. 11175042 and the Fundamental Research Funds for the Central Universities under Grant No. N140505002 and No. N140506002.

\end{document}